\documentclass[preprint,superscriptaddress]{revtex4}
\usepackage{graphicx}
\usepackage{bm}
\usepackage{amssymb}
\usepackage{amsmath}
\begin{document}

\title{Interaction-induced localization of mobile impurities in ultracold systems}

\author{Jian Li}
\affiliation{Texas Center for Superconductivity and Department of Physics, University of Houston, Houston,
Texas 77204, USA}

\author{Jin An}
\affiliation{Texas Center for Superconductivity and Department of Physics, University of Houston, Houston,
Texas 77204, USA}
\affiliation{National Laboratory of Solid State Microstructures and Department of Physics, Nanjing University, Nanjing 210093, China}

\author{C. S. Ting\footnote{email: ting@uh.edu.}}
\affiliation{Texas Center for Superconductivity and Department of Physics, University of Houston, Houston,
Texas 77204, USA}

\date{\today}

\maketitle

\newpage
\textbf{The impurities, introduced intentionally or accidentally into certain materials, can significantly modify their characteristics or reveal their intrinsic physical properties, and thus play an important role in solid-state physics. Different from those static impurities in a solid, the impurities realized in cold atomic systems are naturally mobile. Here we propose an effective theory for treating some unique behaviors exhibited by ultracold mobile impurities. Our theory reveals the interaction-induced transition between the extended and localized impurity states, and also explains the essential features obtained from several previous models in a unified way. Based on our theory, we predict many intriguing phenomena in ultracold systems associated with the extended and localized impurities, including the formation of the impurity-molecules and impurity-lattices. We hope this investigation can open up a new avenue for the future studies on ultracold mobile impurities.}

The experimental studies of impurities in the cold atomic systems\cite{Schirotzek2009, Zipkes2010, Schmid2010, Kohstall2012, Koschorreck2012} has generated a lot of interests in this research area. It provides great opportunities for simulating the static impurity effects which have been shown in solid-state systems, like the pair-breaking effects and in-gap bound states\cite{Vernier2011, Jiang2011, Ohashi2011}. On the other hand, the impurity atoms with mobility possess strikingly unusual effects. This sparks many novel phenomena which are hard to realize in solid materials, such as attractive\cite{Chevy2006, Combescot2007} or repulsive\cite{Cui2010, Massignan2011} Fermi polarons and quantum flutter\cite{Mathy2012}. All these push the study of the impurity effects into new prospects.
Moreover, compared to the systems in real materials, the physical quantities are easier to control with cold atoms. Specifically, the impurity-background (IB) interaction can be precisely tunable in the experiments with the help of an external magnetic field\cite{Kohstall2012, Koschorreck2012}, which facilitates exploring the exotic impurity physics with cold atoms.

The localized impurities, in analogy with the strong coupling polarons in solids\cite{Mahan2000}, were previously studied in several cold atomic systems, including a Bose-Einstein condensate with one or several bosonic impurities\cite{Cucchietti2006, Kalas2006, Sacha2006, Santamore2011, Bruderer2008}, a superfluid Fermi gas with small number of bosonic impurities\cite{Targonska2010} and a Larkin-Ovchinnikov superfluid with fermionic impurities\cite{Li2012}. The extended to localized transition (ELT) of the impurity state is shown to have many outstanding features: (1) Finite value of IB interaction is needed for the localization of the impurity atoms in two and three dimensions (2D and 3D)\cite{Cucchietti2006, Kalas2006, Targonska2010, Li2012}. (2) Any small IB interaction results in the localization of the impurity in one dimension (1D)\cite{Sacha2006, Bruderer2008}. (3) The critical IB interaction for the localization of $N$ ($N>1$) bosonic impurities is smaller than that of a single impurity\cite{Santamore2011}. These features are shared or partly shared in different systems, implying some common behaviors exist in the IB interaction-induced localization of mobile impurities. In this paper, we propose a phenomenological model that is able to explain all the features listed above, and in addition, we predict some exotic features from this model that could be realizable in experiments with ultracold mobile impurities. Although our effective model is proposed for cold atomic systems, it can be extended straightforwardly to other systems with direct IB interactions. Thus our theory provides a general framework for understanding problems associated with the interaction-induced localization of mobile impurities.

\section*{Results}

\subsection*{The effective model}
As shown in the Methods section, the impurities of a total number $N$ immersed in a background with a contact interaction can be effectively described by a universal energy functional regardless of the impurities' statistics:
\begin{eqnarray}
E[\Psi_{1}(\bm{r})...\Psi_{N}(\bm{r})]=\int d\bm{r}[\sum_{i=1}^N\Psi^{*}_{i}(\bm{r})(-\frac{\hbar^{2}}{2m_{I}}\nabla^{2})\Psi_{i}(\bm{r})
-\alpha n_{I}^{2}(\bm{r})+\beta n_{I}^{3}(\bm{r})]
\end{eqnarray}
where $n_{I}(\bm{r})=\sum_{i=1}^N|\Psi_i(\bm{r})|^2$ is the local density of the impurity atoms and $\Psi_i(\bm{r})$ is the wave function of the $i$-th impurity atom. The first term in equation (1) is the kinetic energy and the last two terms are self-induced energies originating in the IB interaction. The energy minimization of equation (1) leads to the one-particle self-consistent equations for the mobile impurities:
\begin{eqnarray}
[-\frac{\hbar^{2}}{2m_{I}}\nabla^{2}-2\alpha n_{I}(\bm{r})+3\beta n_{I}^{2}(\bm{r})]\Psi_{i}(\bm{r})=\mu_{i}\Psi_{i}(\bm{r}),
\end{eqnarray}
where $\mu_i$ are the Lagrange multipliers. The density distributions of the mobile impurities are now determined by an effective system of noninteracting atoms moving in a Kohn-Sham-like\cite{Kohn1965, Ma2012} potential $v_{eff}(\bm{r})=-2\alpha n_{I}(\bm{r})+3\beta n_{I}^{2}(\bm{r})$, and the total energy of the impurities is $E=\sum_{i=1}^N\int d\bm{r}\Psi^{*}_{i}(\bm{r})[\mu_{i}+
\alpha n_{I}(\bm{r})-2\beta n_{I}^{2}(\bm{r})]\Psi_{i}(\bm{r})$.

\subsection*{Localization of single impurity}
Quantum-mechanically, for a single impurity atom confined within a length $\lambda$, its energy in $l-$dimension has a general form: $E\sim\frac{\hbar^2}{2m_I}\frac{l}{\lambda^{2}}-\frac{\alpha}{\lambda^{l}}+\frac{\beta}{\lambda^{2l}}$. Assuming a Gaussian trial wave function $\phi(r)=\frac{1}{(\lambda\sqrt{\pi})^{l/2}}e^{-\frac{r^2}{2\lambda^2}}$, which is similar to the Landau-Pekar treatment\cite{Mahan2000} and was also used in several specific systems\cite{Cucchietti2006, Kalas2006}, the total energy $E$ for a single impurity becomes
\begin{eqnarray}
E=\frac{l\hbar^{2}}{4m_I}\frac{1}{\lambda^{2}}-B_l\frac{\alpha}{\lambda^{l}}+C_l\frac{\beta}{\lambda^{2l}},
\end{eqnarray}
where $B_l=(2\pi)^{-l/2}$ and $C_l=(3\pi^2)^{-l/2}$.
Since thermodynamic stability requires $\alpha$ be positive (see Methods section), the energy contributions from the kinetic part and $\alpha-$part compete with each other. In fact, the kinetic energy favors an extended state with $\lambda^{-1}=0$ and $E=0$, while the $\alpha$-term favors a localized ground state with $\lambda=0$ and $E=-\infty$. A positive $\beta$ then stabilizes the system at a finite $\lambda$ and $E$. The localization here is induced by the impurity itself which creates a local trap in the background through the IB interaction. Below, we show that all the essential physics about the ELT by tuning $\alpha$, including the classification and the critical behaviors of the transition, are manifested in the competition of energy implied in equation (3).

As shown in Fig.1a, a localized state with a finite $\lambda^{-1}$ always gives the minimal energy $E_{min}$ for any positive $\alpha$, indicating that \emph{the impurity is always localized in 1D}. In 2D, $E_{min}$ appears at $\lambda^{-1}=0$ for small $\alpha$, while it appears with a negative value at a finite $\lambda^{-1}$ for large $\alpha$. Therefore there exists a critical value $\alpha_c$ above which the impurity gets localized (see Fig.1b). At the critical point $\alpha=\alpha_c$, $\lambda^{-1}_{c}=0$, indicating that \emph{the ELT is continuous in 2D.}
In 3D as illustrated in Fig.1c, we also have $\alpha_c>0$, but different from 2D, $\lambda^{-1}_{c}>0$ at the critical point $\alpha=\alpha_c$, which suggests \emph{a discontinuous ELT in 3D}. Additionally, there exists another critical value $\alpha_{m}$ in 3D, and in the parameter region $\alpha_{m}<\alpha<\alpha_c$ we have a meta-stable localized state, although the ground state is still extended.

The critical behavior of the ELT in $l$-dimension is then determined by $\frac{\partial E}{\partial \lambda}=0$, and near the transition point we have the optimized localization length and energy:
\begin{eqnarray}
(\lambda^{-1}-\lambda^{-1}_{c})\propto (\alpha-\alpha_{c})^{\nu},\ \ \ E\propto -(\alpha-\alpha_{c})^{\delta}.
\end{eqnarray}
where $\nu=1$, $\delta=2$, $\alpha_c=0$, $\lambda_{c}^{-1}=0$ for 1D; $\nu=1/2$, $\delta=2$, $\alpha_c=constant$, $\lambda_{c}^{-1}=0$ for 2D; $\nu=1$, $\delta=1$, $\alpha_c\propto \beta^{1/4}$, $\lambda_{c}^{-1}\propto \beta^{-1/4}$ in 3D. More precise forms of equation 4 can be found in the Supplementary Information. By solving $\frac{\partial E}{\partial \lambda}=0$ and $\frac{\partial^2 E}{\partial \lambda^2}=0$ simultaneously in 3D, we obtain $\alpha_{m}=0.877\alpha_c$.

The exact impurity profiles and critical behaviors of the ELT can be given by solving equation (2) numerically with the centrosymmetric coordinates of the impurity state. We start with an initial guess of $n_I(r)$, then calculate the potential terms. By minimizing the total energy, we get the $\Psi_i(r)$ and a new $n_I(r)$. This process is repeated until the final density is converged. It also needs to make sure that the integration of $n_I(r)$ over the real space should yield the total number $N$ of the impurities. To extract the universal features from the numerical results, we introduce the length/energy units and dimensionless parameters: $\alpha'$ and $\beta'$ (see the table of the Methods section).
As shown in Fig.2a-c, there always exists a finite localization length in 1D while in 2D and 3D the localization length $\lambda\rightarrow\infty$ for small $\alpha'$, implying that the impurity is always localized in 1D while a critical parameter is needed in 2D and 3D for the localization. From the energy behaviors shown in the inset-plot of Fig.2, we can see that the ELT is continuous in 2D while discontinuous in 3D. Especially, the critical exponent $\delta$ is $1.95$ for 2D and $0.99$ for 3D, and the critical parameters (see Fig.2 and its caption) from the numerical calculations are in good agreement with those from our Gaussian-trial-wave function approaches. This coincides with the conclusions drawn from several specific models, that the Gaussian trial wave function is reliable for the localized impurity state\cite{Cucchietti2006, Kalas2006}.

In our theory, the essential physics can be extracted by expanding the energy to the third order of the impurity density, and taking higher orders into consideration does not change the properties of the ELT as long as they contribute a positive energy. This is similar to the Ginzburg-Landau (G-L) equation where the free energy is expanded to the second order in the density of superconducting order parameter. The localization length $\lambda$, which is in analogy to the coherence length in G-L equations, characterizes the size of the localized wave packet. However, the transition here is between two different states but not phases, and the critical behavior of the transition here is strongly dimension-dependent, which are quite different from G-L theory.

\subsection*{Soliton excitations}
The dynamic of single impurity is described by the time-dependent Schrodinger equation:
\begin{eqnarray}
i\hbar\frac{\partial}{\partial t}\Psi(\bm{r},t)=H_{eff}\Psi(\bm{r},t)
 \end{eqnarray}
where $H_{eff}=-\frac{\hbar^{2}}{2m_{I}}\nabla^{2}-2\alpha |\Psi(\bm{r},t)|^2+3\beta |\Psi(\bm{r},t)|^4$ and we have a series of soliton solutions (detailed derivation is shown in the supplementary information) $\Psi(\bm{r},t)=\Psi(\bm{r}-\bm{v}t)e^{i[\bm{k}\cdot\bm{r}-(\mu+\frac{\hbar^2k^2}{2m_I})t/\hbar]}$, where $\bm{v}=\hbar \bm{k}/m_{I}$. These soliton solutions are characterized by nonconservative momentum parameter $\bm{k}$ and excitation energy $E+\frac{\hbar^2k^2}{2m_I}$, identifying that the localized impurity can be represented by a wave packet of length $\lambda$ moving at a constant speed with the waveform unchanged. In the limit of $\bm{v}\rightarrow0$, the stationary localized impurity wave function $\Psi(\bm{r})$ is fully recovered.

\subsection*{Multi-impurity structures}
Our theory can be applied to $N$($N>1$) indistinguishable ultracold impurities and predict some exotic phenomena. Firstly we consider $N$ noninteracting bosons. Assuming that all the impurities are condensed into a single particle state, i.e., $\Psi_{con}=\sqrt{N}\Psi_1$, we are able to get the critical value of the localization parameter $\alpha_c(N)=\alpha_c/N$ in 2D, and $\alpha_c(N)=\alpha_c/N^{\frac{1}{2}}$ in 3D (details can be found in the supplementary information), where $\alpha_{c}$ is the critical value of $\alpha$ in the single-impurity case, independent of $N$. Therefore, for the non-interacting bosonic impurities, the critical IB interaction for localization is much smaller than single impurity, which is consistent with Ref [18]. For weakly repulsive interacting bosonic impurities, the essential effect of the  impurity-impurity interaction $\lambda_{II}\delta(\bm{r}-\bm{r'})$ is to renormalize the parameter $\alpha$, and the transition boundary (compared to cases with $\lambda_{II}=0$) in all dimensions are shifted to $\alpha_c(N)+\lambda_{II}/2$. Specially, a finite value of IB interaction with $\alpha=\lambda_{II}/2$ is now needed to get the impurity localized in 1D.

Let us now turn our attention to the fermionic impurities. Two localized fermionic impurities are subject to indirect impurity-impurity interactions mediated by the background. This indirect interaction energy can be interpreted as  $E_T-2E_S$, where $E_{S}$ and $E_{T}$ are the energies for one impurity and two correlated impurities, respectively. By constructing a bonding and anti-bonding states of two localized wave functions(available in Supplementary Information) with distance between these two localized impurities $a$ as the variational parameter, and considering $E_T-2E_S$ as function of $a$, transition between bonding and non-bonding states is found. As we can see from Fig.3a, there is a critical value of the dimensionless parameter in 1D and 2D, above which two localized impurity can bond together at an equilibrium inter-impurity distance ${a=a_B}$ with the bonding energy $E_{bond}<0$. In this case the total energy is lowered when two localized impurities bond together to share the same distortions of the background, and the bound pair can be viewed as an impurity bipolaron. Below the critical value two impurities are non-bonding and shortly repulsive to each other. In 3D bipolaron is always formed as long as the impurities are localized. The two-impurity behaviors in different dimensions, including the boundary between bipolaron and non-bonding state, and behaviors of $a_B$ and $E_{bond}$, are summarized in Fig.3b.

Different two-impurity behaviors lead to distinct physical configurations for multi-fermionic impurities. To see this we solve equation (2) self-consistently in a 2D lattice for $N$($N>1$) fermionic impurities (details available at the Supplementary Information). For a strong IB interaction, while two impurities form a bipolaron, as discussed above, impurities with larger numbers are weakly attractive to each other and form a large impurity molecule (see Fig.4a). The weak attraction between impurities can also be reflected in the behaviors of the indirect impurity-impurity interaction energy per impurity, which is found to get more attractive (or negative) with $N$. For a weak IB interaction, the impurities are weakly localized and the impurities are repulsive to each other. As shown in Fig.4b, for a medium impurity density, the impurities form a triangular lattice structure, while the impurities become randomly distributed for small $N$ and the lattice structure collapses for very large $N$ because the size of the system can not accommodate so many localized impurities. Here the indirect impurity-impurity interaction energy per impurity gets more repulsive as $N$ increases. 

As shown in the case with $N=60$ in Fig. 4b, the total kinetic energy of the impurities is quenched and the short-ranged impurity-impurity interaction becomes dominant and repulsive, which leads to the formation of an impurity lattice. This process is similar to the Wigner crystallization in solid-state physics. While the Wigner lattice is formed by electrons with long-range Coulomb interactions, the lattice of neutral impurity atoms is due to the IB interaction.

\section*{Discussion}
We proposed an effective theory for the interaction-induced localization of the ultracold mobile impurities. Beyond its theoretical significance in describing the essential features of the dimension-dependent ELT of the impurity states, our theory also predicted some exotic behaviors of the localized impurities in cold atomic systems, such as molecules and lattice structures, which extends the potential applications of the cold atoms as a quantum simulator for solid-state materials. On the other hand, these features are direct consequences of the impurity effects under the feedback of the background. This is quite different from solid state physics in which the impurity atoms are hardly affected by the background, and marks the unique nature of the mobile impurities. To realize the extended-localized transition and its related phenomena, one may use two-component $^{40}$K superfluid as the background and $^{6}$Li as the impurity atom in 2D, since both the $^{40}$K superfluid\cite{Regal2004} and the three-component $^{40}$K-$^{6}$Li mixture\cite{Spiegelhalder2009} are experimentally accessible. Due to the existence of an energy gap in the spectrum of the background system, the quantum fluctuations and the gapless particle-hole excitations are effectively suppressed, which benefits for the applicability of the mean-field treatment even in low dimensions. As shown in the Supplementary Information, the transition can be achieved by tuning up the scattering length between the impurity and the background to a critical value.

\section*{Methods}
\paragraph*{\bf Derivation of the effective model.}
To derive model (1), we consider a generic Hamiltonian constructed with three parts:
\begin{eqnarray}
H=\int d \bm{r}(h_{I}+h_{B}+h_{IB}),
\end{eqnarray}
where $h_{I}$ and $h_{B}$ are local functions of the coordinate $\bm{r}$. $h_{I}$ is the density of the impurity kinetic energy. $h_{B}$ contains the kinetic energy density of the background atoms, and the interaction term between the background atoms, which varies for different interacting systems. The IB interaction part is
\begin{eqnarray}
h_{IB}=U_{IB}\hat{n}_{I}(\bm{r})\hat{n}_{B}(\bm{r}),
\end{eqnarray}
and it can be rewritten as
\begin{eqnarray}
h_{IB}=U_{IB}\langle\hat{n}_{I}(\bm{r})\rangle\hat{n}_{B}(\bm{r})+U_{IB}\langle\hat{n}_{B}(\bm{r})\rangle\hat{n}_{I}(\bm{r})
-U_{IB}\langle\hat{n}_{I}(\bm{r})\rangle\langle\hat{n}_{B}(\bm{r})\rangle
\end{eqnarray}
in the mean-field level.
Within the local density approximation, the local energy density is then given by
\begin{eqnarray}
E(\textbf{r})=E_{B}(\bm{r})|_{\mu_{B}\rightarrow\mu_{B}-U_{IB}n_{I}(\bm{r})}+E_{I}(\bm{r}),
\end{eqnarray}
where $n_{I,B}=\langle \hat{n}_{I,B}(\bm{r})\rangle$, $E_B|_{\mu_{B}}=\langle h_{B}\rangle$, $E_I=\langle h_{I}\rangle$, and $\mu_B$ is the chemical potential of the background. By expanding the background energy $E_{B}$ we get
\begin{eqnarray}
E(\bm{r})=E_{I}(\bm{r})+E_{B}(\bm{r})+\sum_{m}C_{m}[-U_{IB} n_{I}(r)]^m,
\end{eqnarray}
 where $C_{m}=\frac{\partial^m E_{B}}{m!\partial\mu^m}|_{\mu=\mu_{B}}$. If the impurity concentration is dilute, i.e., $N/N_B\ll1$, where $N_B$ is the total number of the background atoms, the impurities only induce slight changes to the background. In this case it is sufficient to keep $n_{I}(\bm{r})$ to the third order, and get the total energy $E=\int d\bm{r} E(\bm{r})$ as presented in equation (1) by neglecting impurity-irrelevant terms. The parameters are then given by
\begin{eqnarray}
\alpha&=&-\frac{\partial^2 E_{B}}{2\partial\mu^2}|_{\mu=\mu_{B}}U_{IB}^2=\frac{\partial n_{B}}{2\partial\mu}|_{\mu=\mu_{B}}U_{IB}^2\\ \beta&=&-\frac{\partial^3 E_{B}}{6\partial\mu^3}|_{\mu=\mu_{B}}U_{IB}^3=\frac{\partial^2 n_B}{6\partial\mu^2}|_{\mu=\mu_{B}}U_{IB}^3.
\end{eqnarray}
Clearly for a homogeneous background in the absence of impurities, $\alpha,\beta$ are $\bm{r}$-independent. For most systems the thermodynamical stability requires a positive $\alpha$, while the sign of $\beta$ is largely relevant to the IB interaction and the details of the background.

\paragraph*{\bf Length and energy unit.}
In 1D we define the length unit $a_0$ and the energy unit $E_0$ through $\frac{\hbar^2}{2m_Ia_{0}^{2}}\sim\frac{\alpha}{a_{0}}$, and in 2D and 3D they are defined through $\frac{\hbar^2}{2m_Ia_{0}^{2}}\sim\frac{\beta}{a_{0}^{2l}}$ where $l=2,3$. Then we get the length and energy unit in the table below. Accordingly, the parameters $\alpha,\beta$ have the dimensionless forms $\alpha',\beta'$ which are also summarized in the table below. Notice there is only one independent dimensionless parameter left in each dimension.
\begin{center}
\begin{tabular}{|c|c|c|c|c|}
\hline
 \ & $a_{0}$ & $E_{0}$ & $\alpha'$ & $\beta'$\tabularnewline
\hline
1D & $\frac{\hbar^{2}}{2m_I\alpha}$ & $\frac{2m_I\alpha^{2}}{\hbar^{2}}$ & $1$ & $\frac{2m_I\beta}{\hbar^{2}}$\tabularnewline
\hline
2D & $(\frac{2m_I\beta}{\hbar^{2}})^{\frac{1}{2}}$ & $(\frac{\hbar^{2}}{2m_I\sqrt{\beta}})^{2}$ & $\frac{2m_I\alpha}{\hbar^{2}}$ & $1$\tabularnewline
\hline
3D & $(\frac{2m_I\beta}{\hbar^{2}})^{\frac{1}{4}}$
& $(\frac{\hbar^{2}}{2m_I\sqrt[3]{\beta}})^{\frac{3}{2}}$ & $\alpha(\frac{2m_I}{\sqrt[3]{\beta}\hbar^{2}})^{\frac{3}{4}}$ & $1$\tabularnewline
\hline
\end{tabular}
\end{center}

\section*{Acknowledgments}
This work was supported by the Texas Center for Superconductivity at the University of Houston and by the Robert A. Welch Foundation under Grant No. E-1146. Jin An was also supported by NSFC(China) Project No.1117416.

\section*{Author contributions}
All the authors contributed equally to this work.
All the authors worked closely to propose this study.
J.L. and J.A. performed the calculations.
J.L., J.A. and C.S.T. analyzed the results and wrote the manuscript.

\section*{Additional information}
The authors declare no competing financial interests.

\clearpage \thispagestyle{empty}
\begin{figure}[t]
\includegraphics[width=5.4in] {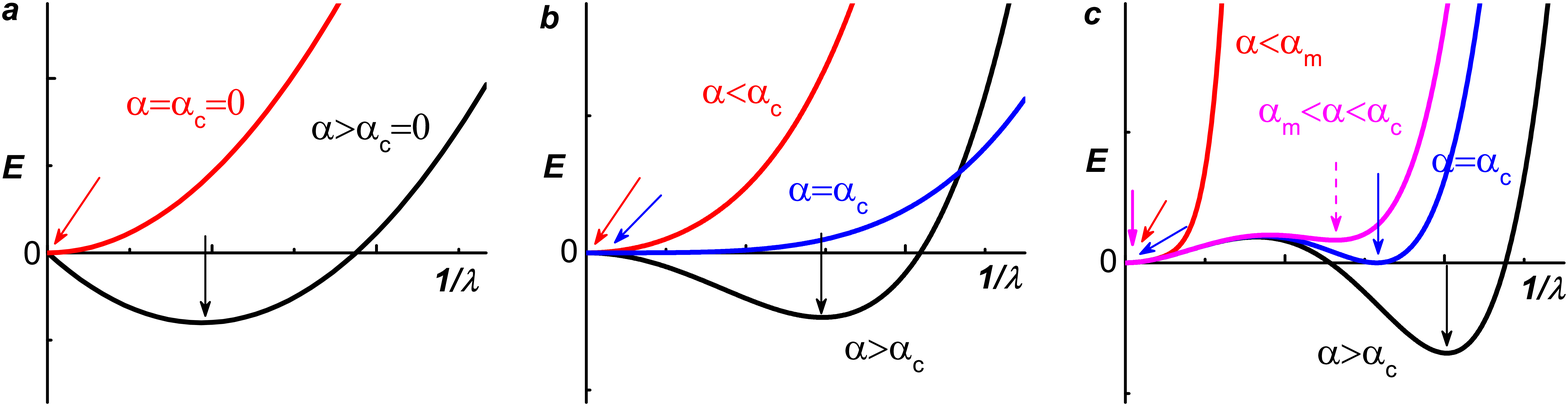}
\caption{\textbf{Schematic diagram for single impurity energy as function of $\lambda^{-1}$ with $\mathbf{\beta>0}$.} \textbf{a}-\textbf{c}, 1D, 2D and 3D. The arrows in solid lines indicate the global minimums of the energies, while the arrow in dashed line in (\textbf{c}) marks a local minimum of $E$.}
\label{Fig1}
\end{figure}

\clearpage \thispagestyle{empty}
\begin{figure*}
\includegraphics[width=6.3in] {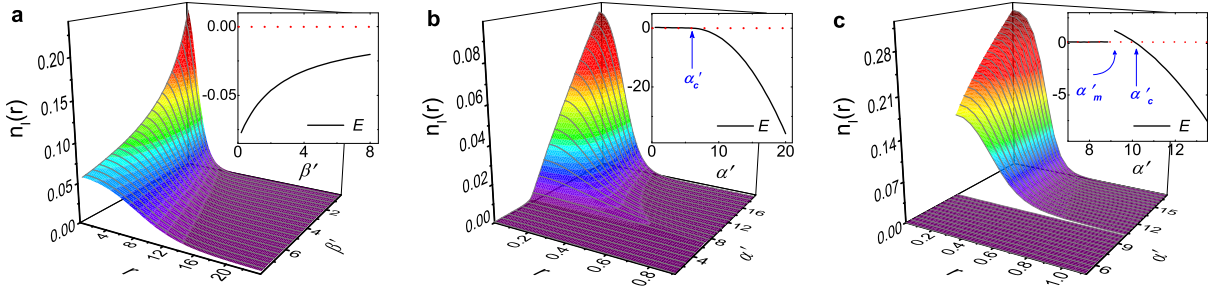}
\caption{\textbf{The impurity structure as function of the dimensionless parameter.} \textbf{a}-\textbf{c}, 1D, 2D and 3D. The asymptotic behavior of the localized wave function indicated by equation (2) is parabolic at $r\rightarrow0$: $\Psi(r)\rightarrow
\Psi(0)+\frac{1}{2}\Psi''(0)r^{2}$, and exponential at $r\rightarrow\infty$: $\Psi(r)\rightarrow e^{-kr}$.
Inset: The corresponding energy of the impurity as function of the dimensionless parameter in (\textbf{a}-\textbf{c}). The dashed (red) lines represent the energy of the extended states. The blue arrows mark the critical values of the dimensionless parameters for the occurrence of the localized or meta-stable localized state. The critical values given by the numerical results are (2D) $\alpha'_c=6.0$; (3D) $\alpha'_c=10.28$ and $\alpha'_{m}=0.892\alpha'_c$, which are comparable with the results from Gaussian trial wave function method where (2D) $\alpha'_c=2\pi$; (3D) $\alpha'_c=10.51$ and $\alpha'_{m}\approx0.877\alpha'_c$. Here $r$ and $E$ is in the unit of $a_0$ and $E_0$, respectively.}
\end{figure*}

\clearpage \thispagestyle{empty}
\begin{figure}[tb]
\includegraphics[width=5.4in] {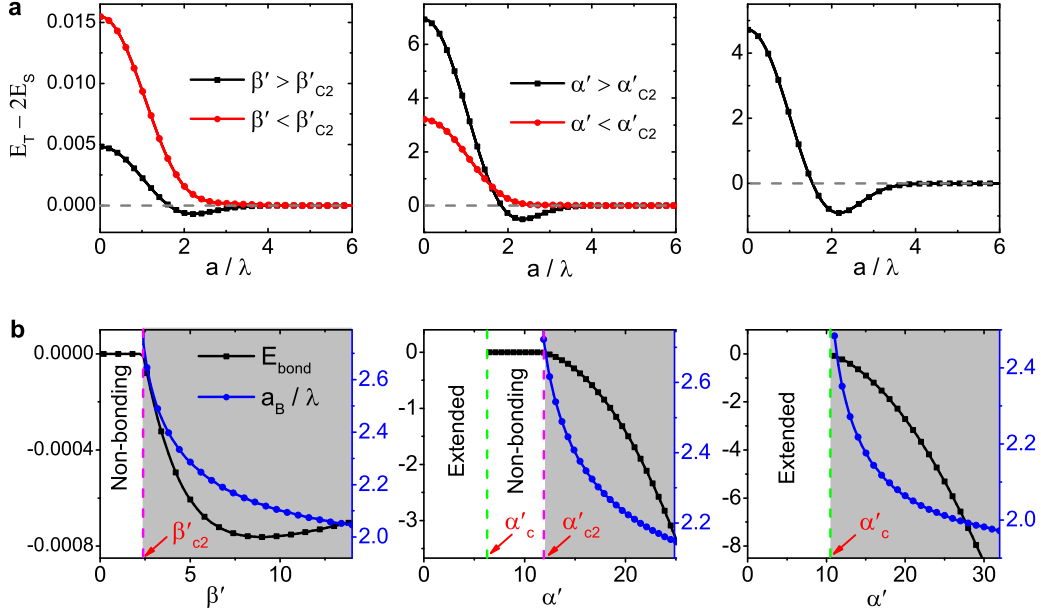}
\caption{\textbf{Two-impurity bonding energy and bonding length.} \textbf{a}, the background mediated impurity-impurity energy $E_T-2E_S$ for two localized ferminoic impurities as function of inter-impurity distance in 1D, 2D and 3D (from left to right). \textbf{b}, the bonding energy $E_{bond}$ and the bonding length $a_B$ as function of dimensionless parameters in 1D, 2D and 3D (from left to right). Notice here $E_{bond}$ is defined as the minimum $E_T-2E_S$ and $a_B$ is the equilibrium inter-impurity distance. The shadow region marks the parameter space of the formation of the bipolaron. For non-bonding states we have $E_{bond}=0$ and the absence of $a_B$. The numerical results give the critical value $\beta'_{c2}=2.4$ at 1D and $\alpha'_{c2}=11.9$ at 2D. $a_B/\lambda$ decreases when the dimensionless parameters are increased in all dimensions, while $E_{bond}$ shows a non-monotonic behavior in 1D, and at very large $\beta'$, $E_{bond}$ approaches zero but still keeps negative, indicating that the bonding state is very weak in this case. }
\label{Fig3}
\end{figure}

\clearpage \thispagestyle{empty}
\begin{figure}[tb]
\includegraphics[width=4.2in] {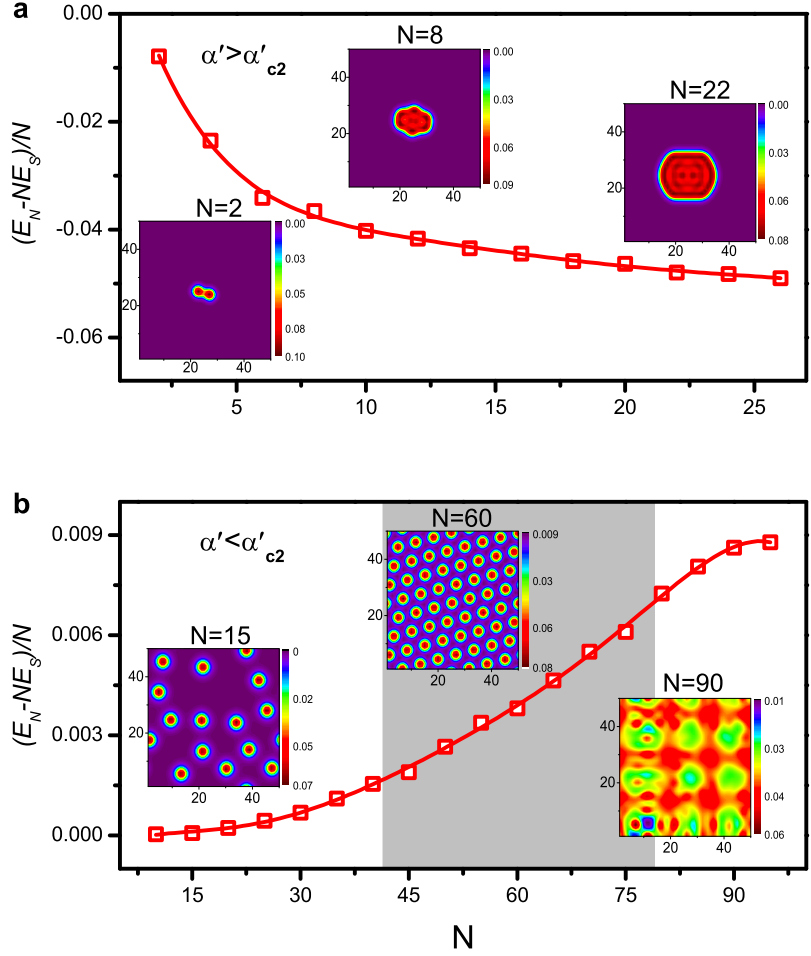}
\caption{\textbf{Multi-impurity structure and indirect impurity-impurity interaction energy.} \textbf{a}-\textbf{b}, plot of indirect impurity-impurity interaction energy per impurity $(E_N-NE_S)/N$ (red solid lines and signals) as function of $N$ in a 2D lattice model with dimensionless parameter (\textbf{a}) $\alpha'=15.0$, $\beta'=1.0$, (\textbf{b})$\alpha'=7.0$, $\beta'=1.0$. Here $E_{N,(S)}$ is the total energy for $N$ (single) fermionic impurities, and we use the length and energy unit defined in the table of the Methods section. Inset: contour-plot of the impurity density distribution for (\textbf{a}) $N=2,8,22$ and (\textbf{b}) $N=15,60,90$. The shadow region in (\textbf{b}) marks where a stable lattice appears. }
\label{Fig4}
\end{figure}

\end{document}